\documentclass[11pt]{article}
\usepackage{amsmath,amsthm,amscd,amssymb}
\usepackage{amsfonts}
\usepackage{amsmath}
\usepackage[english]{babel}
\newtheorem{prop}{Proposition}[section]

\newtheorem{theo}[prop]{Theorem}
\newtheorem{lem}[prop]{Lemma}

\newtheorem{rem}[prop]{Remark}

\def\1{1\!{\rm l}}

\begin{document}
\title{About a result of S.M. Kozlov}\maketitle
\begin{center}
{\bf Hatem NAJAR}\\
\footnotesize{D\'epartement de Math\'ematiques Physiques I.P.E.I.
Monastir, 5000 Monastir Tunisie . } \footnote{Researches partially
supported by  CMCU N 02/F1511 and N 04/S1404 projects. }
\end{center}
\begin{abstract}
\noindent We give an alternative proof and improve upon a result
of S.M.~Kozlov \cite{ko}. It deals with the asymptotic of the
integrated density of states of the acoustic operator
$\displaystyle H_\omega=-\nabla\rho_\omega\nabla$, at the bottom
of the spectrum.
\end{abstract}
{
{\small\sf 2000 Mathematics Subject Classification :81Q10, 35P05,
37A30,
47F05.}}

\section{Introduction}
\label{intro} Let $H_{\omega}$, be the self adjoint operator on
$L^2({\Bbb{R}}^d)$ formally defined by:
\begin{equation}
H_{\omega}=H(\rho _{\omega})=-\nabla \cdot\rho
_{\omega}\cdot\nabla, \label{b1}
\end{equation}
where $\rho _{\omega} $ is a positive and bounded
function.\newline $H_{\omega}$ is called the acoustic operator,
see \cite{4} for the physical interpretations.
\newline Let us start by defining the main object of our study:
the integrated density of states.  For this, we consider $\Lambda$
a cube of ${\ \Bbb{R}}^d$. We note by $H_{\omega, \Lambda}$ the
restriction of $H_{\omega}$ to $\Lambda$ with self-adjoint
boundary conditions. As $H_{\omega}$ is elliptic, the resolvent of
$H_{\omega, \Lambda}$ is compact and, consequently, the spectrum
of $H_{\omega, \Lambda} $ is discrete and is made of isolated
eigenvalues of finite multiplicity \cite{21}. We define
\begin{equation}
N_{\Lambda}(E)=\frac{1}{\mathrm{vol}(\Lambda)}\cdot
\#\{\mathrm{{eigenvalues \ of \ } A_{\omega,\Lambda}\leq E\}.}
\label{aha}
\end{equation}
Here $\mathrm{{vol}(\Lambda)}$ is the volume of $\Lambda$ in the
Lebesgue sense and $\# E$ is the  cardinal {of $E$.}\newline
It is shown that the limit of $N_{\Lambda}(E)$ when $\Lambda $ tends to $%
\Bbb{R}^d$ exists almost surely and is independent of the boundary
conditions. It is called the \textbf{integrated density of states} of $%
A_{\omega}$ (IDS as acronym). See \cite{Pa-Fi:92}.\newline The
question we are interested in here regards the behavior of $N$ at
the bottom of the spectrum of $H_{\omega}$. In previous works
\cite{W0,W1,W2,W3}, the author gives the behavior of $N$ at the
internal band edges of the spectrum of (\ref{b1}). It was a
Lifshitz behavior ($N$ decreases exponentially fast). In the
present situation, for the bottom of the spectrum, it is known
that it can't decrease more than polynomially fast, \cite{ko}.
Here we compare the behavior of $N$ to the behavior of the IDS of
some periodic operator with exponentially precision.
\newline \textit{$\mathbf{Acknowledgements.}$ The author would
like to thank professor Fr\'ederic KLOPP for interesting discution
concerning this work and professor Mabrouk Ben Ammar for many
helpful.}
\subsection{The
model } Consider the random Schr{\"o}dinger operator
\begin{equation}
  \label{eq:1}
  H_\omega=-\nabla\frac{1}{\rho_\omega}\nabla.
\end{equation}
Where $\rho_\omega$ is a bounded, $\mathbb{Z}^d$-ergodic random
field such that there exists some constant $\rho_*>1$, satisfying
\begin{equation}
  \label{eq:3}
  \rho_*\leq\rho_\omega \leq\rho_*.
\end{equation}
We assume that $\rho_\omega$ is of Anderson type i.e. it has the
form
\begin{equation}
  \label{eq:6}
  \rho_\omega(x)=\rho^+(x)+\sum_{\gamma\in\mathbb{Z}^d}\omega_\gamma
  \rho^0(x-\gamma)
\end{equation}
where
\begin{itemize}
\item $\rho^+$ is a-$\mathbb{Z}^d$-periodic
  measurable function,
\item $\rho^0$ is a  compactly supported
  measurable function,
\item $(\omega_\gamma)_{\gamma\in \mathbb{Z}^d}$ are non trivial, i.i.d. random
  variables.
  \end{itemize}
The choice of our model ensures that $A_{\omega}$ is a measurable
family of self-adjoint operators and ergodic \cite{6,Pa-Fi:92}.
Indeed, if $\tau_{\gamma}$
refers to the translation by $\gamma$, then $(\tau_{\gamma})_{\gamma\in {%
\Bbb{Z}^d}}$ is a group of unitary operators on
$L^{2}({\Bbb{R}}^d)$ and for ${\gamma\in {\Bbb{Z}^d}}$ we have
$$
\tau_{\gamma}A_ {\omega}\tau_{-\gamma}=A_{\tau_{\gamma}\omega}.
$$
According to \cite{6,Pa-Fi:92}, we know that there exists $\Sigma,
\Sigma _ {pp}, \Sigma _ {ac}$ and $\Sigma _ {sc}$ closed and
non-random sets of ${\Bbb{R}}$ such that $\Sigma$ is the spectrum
of $A_{\omega}$ with probability one and such that if
$\sigma_{pp}$ (respectively $\sigma_{ac}$ and $\sigma_{sc}$)
design the pure point spectrum (respectively the absolutely
continuous and singular continuous spectrum) of $A_{\omega}$, then
$\Sigma _ {pp}=\sigma _ {pp}, \Sigma _ {ac}=\sigma _ {ac}$ and
$\Sigma _ {sc}=\sigma _ {sc}$ with probability one.
 $H_\omega$ is defined as the Friedrichs extension of the following positive
quadratic form
$$\mathcal{H}_{\omega}[\psi]=\langle \rho_\omega\nabla\psi,\nabla
\psi\rangle\ \ ,\  \psi \in H^1_{\text{loc}}(\mathbb{R}^d).$$
\subsection{The result}We shall
prove
\begin{theo}
  \label{thr:1} There exists $\alpha,\ \tau >0$ and $C>1$ such that when $E\to 0^+$
  we
  have,
\begin{equation}
    \label{eq:2}
    \overline{\mathcal{N}}(E-E^{\alpha})-Ce^{-E^{-\tau}}\leq \mathcal{N}(E)\leq
    \overline{\mathcal{N}}(E+E^{\alpha})+Ce^{-E^{-\tau}}
  \end{equation}
  where $\overline{\mathcal{N}}$ is the integrated density of states of
  the following periodic operator
  \begin{equation}
    \label{eq:4}
    \overline{H}=-\nabla\overline{\rho}\nabla;
  \end{equation}
  and $\overline{\rho}=\mathbb{E}(\rho_{\omega})$.
\end{theo}
\begin{rem} 1)The improvement over Kozlov's result essentially consists in
the estimate of the remainder term and the exponential precision .
$$\mathcal{N}(E)\to \overline{\mathcal{N}}(E+o(E))\ \ \rm{exponentially\ as }\ E\to 0^+.$$
2)We don't believe this estimate to be optimal: namely, we expect
the exponent $\alpha$ to be larger than the one we found in the
present study.
\end{rem}
\section{Proof of Theorem~\ref{thr:1}}
\subsection{The periodic approximation}
\vskip.2cm\noindent Pick $n\in\mathbb{N}\setminus\{0\}$ and define
the following periodic Schr{\"o}dinger operator
\begin{equation*}
  H^n_{\omega}=-\nabla\rho^n_{\omega}\nabla.
\end{equation*}
Here, $$\rho_\omega^n=\rho^{+,n}+\rho_{\omega}^{0,n}=\rho^+(x)+\sum_{\gamma \in \Lambda_{n}\cap {\mathbb{Z}^{d}}%
}\omega _{\gamma }\sum_{\beta \in
(2n+1){\mathbb{Z}^{d}}}\rho^0(x-\gamma+\beta).$$ Where
$\Lambda_{k}$ is the cube
$$
\Lambda_{n}=\{x\in {\mathbb{R}}^{d};\forall 1\leq j\leq d,\ -\frac{2k+1}{2}%
<x_{j}\leq \frac{2k+1}{2}\}.$$For $\omega$ fixed and
$n\in\mathbb{N}^*$, $H^n_{\omega}$ is a
$(2n+1)\mathbb{Z}^d$-periodic self-adjoint Schr{\"o}dinger
operator. \\ Let $\overline{\omega}=\mathbb{E}(\omega_0)$ and
$\displaystyle \overline{\rho^{0,n}}=\sum_{\gamma \in \Lambda_{n}\cap {\mathbb{Z}^{d}}%
}\overline{\omega}\sum_{\beta \in
(2n+1){\mathbb{Z}^{d}}}\rho^0(x-\gamma+\beta)$.
\subsection{Some Floquet Theory} Now we review some standard
facts from the Floquet theory for periodic operators. Basic
references of this material are in \cite{21}.\newline Let the
torus $\mathbb{T}^*_{2n+1}=\mathbb{R}^d/2\pi(2n+1)\mathbb{Z}^d$.
We define ${ {\mathcal{H}}}_n$ by \begin{multline*}
{{\mathcal{H}}}_n=\{u(x,\theta )\in
L_{loc}^{2}({\mathbb{R}}^{d})\otimes L^{2}(
{\mathbb{T}}_{2n+1}^{\ast });\\ \forall (x,\theta ,\gamma )\in
\mathbb{R}^{d}\times \mathbb{T}^{\ast }_{2n+1}\times
(2n+1)\mathbb{Z}^{d};\ u(x+\gamma ,\theta )=e^{i\gamma \theta
}u(x,\theta )\}.
\end{multline*}
There exists $U$ a unitary isometry from $L^{2}({\mathbb{R}}^{d})$ to ${{%
\mathcal{H}}}_n$ such that $H_{\omega}^n$ admits the following
Floquet decomposition \cite{21}
\begin{equation*}
UH_{\omega}^nU^{\ast }=\int_{{\mathbb{T}_{2n+1}^{\ast }}}^{\oplus
}H_{\omega}^n(\theta)d\theta .
\end{equation*}
Here $H_{\omega}^n(\theta )$ is the self adjoint operator on
${{\mathcal{H}}} _{n,\theta }$ defined as the operator
$H_{\omega}^n$ acting on ${{\mathcal{H}}}^1 _{n,\theta }$ with
$$
{\mathcal{H}}_{n,\theta }=\{u\in
L_{loc}^{2}(\mathbb{R}^{d});\forall \gamma \in
(2n+1){\mathbb{Z}^{d}},u(x+\gamma )=e^{i\gamma \theta }u(x)\},$$
and $$ {{\mathcal{H}}}^1 _{n,\theta }=\{u\in
{{\mathcal{H}}}_{n,\theta };\
\partial_{x}^{\alpha}u\in {{\mathcal{H}}} _{n,\theta };\
|\alpha|=1\}.
$$
As $H_{\omega}^n$ is elliptic, we know that, $H_{\omega}^n(\theta
)$ has a compact resolvent; hence its spectrum is discrete
\cite{21}. We denote its eigenvalues, called Floquet eigenvalues
of $H_{\omega}^n(\theta)$, by
\begin{equation*}
E_{0}(n,\omega,\theta )\leq E_{1}(n,\omega,\theta )\leq \cdot
\cdot \cdot \leq E_{k}(n,\omega,\theta )\leq \cdot \cdot \cdot .
\end{equation*}
The corresponding eigenfunctions are denoted by $(w(x,\cdot )_{k})_{k\in {%
\mathbb{N}}}$. The functions $(\theta \rightarrow E_{k}(n,\omega,\theta ))_{k\in {%
\mathbb{N}}}$ are Lipshitz-continuous, and we have
\begin{equation*}
E_{k}(n,\omega,\theta )\rightarrow +\infty \ \ \mathrm{as}\
k\rightarrow +\infty \ \ \mathrm{uniformly\ in}\ \ \theta .
\end{equation*}
The spectrum $\sigma (A_{\omega}^n)$ of $A_{\omega}^n$ has a band
structure. (i.e $\displaystyle \sigma (A_{\omega}^n)=\bigcup
_{k\in \mathbb{N}}E_{k}(n,\omega,{\mathbb{T}^{\ast }})$).\newline
Let $\mathcal{N}^n_{\omega}$ be the integrated density of states
of $H_{\omega}^n$; it satisfies
\begin{equation}
  \label{idosper}
  \mathcal{N}^n_{\omega}(E)=\sum_{k\in\mathbb{N}}\frac{1}{(2\pi)^d}
  \int_{\{\theta\in\mathbb{T}^*_{2n+1};\ E_k(n,\omega,\theta)\leq
  E\}}d\theta =\frac{1}{(2\pi)^d}\int_{\mathbb{T}^*_{2n+1}}
  \mathcal{V}(H_{\omega}^n(\theta),E)d\theta.
\end{equation}
Here $\mathcal{V}(B,E)$ is the number of eigenvalues of $B$ less
or equal to $E$. Let $d\mathcal{N}_{\omega}^n$ be the derivative
of $\mathcal{N}_{\omega}^n$ in the distribution sense. As
$\mathcal{N}_{\omega}^n$ is increasing, $d\mathcal{N}_{\omega}^n$
is a positive measure; it is the density of states of
$H_{\omega}^n$. We denote by $d\mathcal{N}$
the density of states of $H_{\omega }$. For all $\varphi \in C_{0}^{\infty }(%
{\mathbb{R}}),d\mathcal{N}_{\omega}^n$ verifies \cite{kp8},
\begin{eqnarray}
\langle \varphi ,d\mathcal{N}_{\omega}^n\rangle &=&\frac{1}{(2\pi
)^{d}}\int_{\theta \in \mathbb{T}_{n}^{\ast
}}\mathrm{tr}_{{\mathcal{H}}_{\theta }}\Big(\varphi
(H_{\omega }^n(\theta ))\Big)d\theta , \notag \\
&=&\frac{1}{\mathrm{vol}(C_{k})}\mathrm{tr}\Big(\chi
_{C_{k}}\varphi (H_{\omega}^n)\chi _{C_{k}}\Big), \label{1sla}
\end{eqnarray}
where for $\Lambda \subset {\mathbb{R}}^{d},\ \chi _{\Lambda }$
will design the characteristic function of $\Lambda $ and
$\mathrm{tr}(A)$ is the trace
of $A$ (we index by ${\mathcal{H}}_{\theta }$ if the trace is taken in ${%
\mathcal{H}}_{\theta }$).
\begin{lem} [\cite{kp8}]\label{T20}
For any $\varphi\in C_{0}^{\infty}({\Bbb{R}})$ and for almost all
$\omega\in \Omega $ we have
$$\lim_{n\rightarrow \infty
}\mathbb{E}(\langle
\varphi,d\mathcal{N}_{\omega}^n\rangle)=\langle
\varphi,d\mathcal{N}\rangle.$$
\end{lem}
Moreover, we have that the IDS of $H_{\omega}$ is exponentially
well-approximated by the expectation of the IDS of the periodic
operators $H_{\omega}^n$ when $n$ is polynomial in
$\varepsilon^{-1}$. More precisely we have
\begin{theo}[\cite{kp8}]
  \label{thr:5}
  Pick $\eta>0$ and $I\subset\mathbb{R}$, a compact interval. There exists
  $\varepsilon_0>0$ and $\rho>0$ such that, for $E\in I$, $\varepsilon\in(0,\varepsilon_0)$
  and $n\geq\varepsilon^{-\rho}$, one has
  \begin{multline}
    \label{estpre}
    \mathbb{E}(\mathcal {N}^n_{\omega}(E+\varepsilon/2))- \mathbb{E}(\mathcal
      {N}^n_{\omega}(E-\varepsilon/2))- e^{-\varepsilon^{-\eta}}\\
    \leq{\mathcal {N}}(E+\varepsilon)-\mathcal {N}(E-\varepsilon)\leq
    \\ \mathbb{E}({\mathcal N}^n_{\omega}(E+2\varepsilon))-\mathbb{E}(\mathcal
      {N}^n_{\omega}(E-2\varepsilon))+e^{-\varepsilon^{-\eta}}.
  \end{multline}
\end{theo}
\begin{rem}
This lemma is proven in \cite{kp8} for the Schr\"{o}dinger case.
It is still true for our case. The proof is based on the
Helffer-Sj\"{o}strand formula and the resolvent equation with the
exponential decay of the resolvent kernels (the Combes-Thomas
argument).
\end{rem}
Now we study the periodic approximations. For a vector space $E$,
we note by $dim(E)$ the dimension of $E$.  We have,
\begin{eqnarray*}
{\mathcal{V}}(H_{\omega}^n(\theta),E)&=&\sup
dim\{\mathcal{E}\subset \mathcal{H}_{n,\theta}^n, \ {\rm{such\
that}},\forall u\in \mathcal{E};\ \langle H_{\omega}^{n}(\theta)
u,u\rangle
\leq E\|u\|^2\}\\
&=&\sup dim\{\mathcal{E}\subset \mathcal{H}_{n,\theta}^n, \
{\rm{such\ that}},\forall u\in \mathcal{E};\\
 && \langle
\Big(H_{\omega}^{n}(\theta)-\overline{H}^n(\theta)\Big) u,u\rangle
+\langle \overline{H}^n(\theta)u,u\rangle \leq E\|u\|^2\}.
\end{eqnarray*}
Let, $\alpha>0$ and
$$
\mathcal{E}_{1}^{\alpha}(\theta)=\{u\in \mathcal{H}_{n,\theta}^1;
\ \Big|\langle
\Big(H_{\omega}^{n}(\theta)-\overline{H}^n(\theta)\Big) u,u\rangle
\Big|\leq E^{\alpha} \|u\|^2\},
$$
and
$$
\mathcal{E}_{2}^{\alpha}(\theta)=\{u\in \mathcal{H}_{n,\theta}^1;
\ \Big|\langle \Big(H_{\omega}^{n}(\theta)-\overline{H}^n(\theta)
\Big)u,u\rangle \Big|\geq E^{\alpha} \|u\|^2\}.
$$
Then we have
\begin{eqnarray*}
{\mathcal{V}}(H_{\omega}^n(\theta),E)&\leq&\sup
dim\{\mathcal{E}\subset \mathcal{E}_{1}^{\alpha}(\theta), \
{\rm{such\ that}},\forall u\in \mathcal{E};\ \langle
H_{\omega}^{n}(\theta) u,u\rangle \leq
E\|u\|^2\}\\
&+&\sup dim\{\mathcal{E}\subset \mathcal{E}_{2}^{\alpha}(\theta),
\ {\rm{such\ that}},\forall u\in \mathcal{E};\ \langle
H_{\omega}^{n}(\theta)
u,u\rangle \leq E\|u\|^2\}\\
&\leq&\sup dim\{\mathcal{E}\subset
\mathcal{E}_{1}^{\alpha}(\theta), \ {\rm{such\ that}},\forall u\in
\mathcal{E};\ \langle
\overline{H}^{n}(\theta)u,u\rangle \leq (E+E^{\alpha})\|u\|^2\}\\
&+&\sup dim\{\mathcal{E}\subset \mathcal{E}_{2}^{\alpha}(\theta),
\ {\rm{such\ that}},\forall u\in \mathcal{E};\langle
H_{\omega}^{n}(\theta) u,u\rangle \leq E\|u\|^2\}.
\end{eqnarray*}
So, we get
\begin{multline}
{\mathcal{V}}(H^n_{\omega}(\theta),E)\leq
{\mathcal{V}}(\overline{H}^n(\theta),(E+E^{\alpha}))\\ +\sup
dim\{\mathcal{E}\subset \mathcal{E}_{2}^{\alpha}(\theta), \
{\rm{such\ that}},\forall u\in \mathcal{E};\langle
H_{\omega}^{n}(\theta) u,u\rangle \leq E\|u\|^2\}.\label{W1}
\end{multline}
Now integrating both sides of (\ref{W1}) over
$\mathbb{T}^*_{2n+1}$ and taking into account (\ref{idosper}), we
get that
\begin{multline}\label{eq:15}
\mathcal{N}_{\omega}^n(E)\leq
\overline{\mathcal{N}}^n((E+E^{\alpha}))+\\
\frac{1}{(2\pi)^d}\int_{\mathbb{T}^*_{2n+1}}\text{dim}\{\mathcal{E}\subset
\mathcal{E}_{2}^{\alpha}(\theta), \ {\rm{such\ that}},\forall u\in
\mathcal{E} ;\ \langle H_{\omega}^{n}(\theta) u,u\rangle \leq
E\|u\|^2\}d\theta.
\end{multline}
Where $\overline{\mathcal{N}}^n$ is the IDS of
$\overline{H}^n$.\newline Notice that $\text{dim}\{
\mathcal{E}\subset \mathcal{E}_{2}^{\alpha}(\theta), \ {\rm{such\
that}},\forall u\in \mathcal{E}; \langle H_{\omega}^{n}(\theta)
u,u\rangle \leq E\|u\|^2\}$  is
 bounded by the number of eigenvalues of $-\Delta^{n}(\theta)$ less
 than $E\rho*$ which is it self bounded by $Cn^d$ ($C$ depends
 only on $E$). As the volume of $\mathbb{T}^*_{2n+1}$ is
 $\displaystyle (2\pi(2n+1))^{-d}$ we get that for some $C>0$ we
 have
 \begin{equation}\label{eq:16}
 \mathbb{E}(\mathcal{N}_{\omega}^n(E))\leq
 \overline{\mathcal{N}}^n\Big((E+E^{\alpha})\Big)+C\mathbb{P}(\Omega_{n,E,\alpha}).
 \end{equation}
 With
 $$
\Omega_{n,E,\alpha}= \Big\{\omega;\ \exists u\in
\mathcal{E}_{2}^{\alpha}(\theta); \|u\|_{L^2(\mathbb{R}^d)}=1;\
\langle \nabla u,u\nabla \rangle \leq E\rho^{*}\|u\|^2\Big\}.
 $$
Now let us consider
 \begin{multline*}
 {\mathcal{V}}(\overline{H}^{n}(\theta),(E-E^{\alpha}))=\\  \sup
 dim\{\mathcal{E}\subset \mathcal{H}^{1}_{n,\theta}, \
{\rm{such\ that}},\forall u\in \mathcal{E};\
 \langle\overline{H}^{n}(\theta)u,u\rangle \leq
 (E-E^{\alpha})\|u\|^2\}.
 \end{multline*}
 \begin{multline*}=\sup dim\{\mathcal{E}\subset \mathcal{H}^{1}_{n,\theta}, \
{\rm{such\ that}},\forall u\in \mathcal{E};\\
 \langle
\Big(\overline{H}^n(\theta)- H_{\omega}^{n}(\theta)\Big)u,u\rangle
+\langle H_{\omega}^{n}(\theta)u,u\rangle \leq
(E-E^{\alpha})\|u\|^2\}
\end{multline*}
\begin{multline*}
\leq \sup dim\{\mathcal{E}\subset
\mathcal{E}_{1}^{\alpha}(\theta), \ {\rm{such\ that}},\forall u\in
\mathcal{E};\ \langle \langle H_{\omega}^n(\theta)u,u\rangle \leq
E\|u\|^2\} \\ +\sup dim\{\mathcal{E}\subset
\mathcal{E}_{2}^{\alpha}(\theta), \ {\rm{such\ that}},\forall u\in
\mathcal{E};\ \langle\overline{H}^{n}(\theta)u,u\rangle \leq
 (E-E^{\alpha})\|u\|^2\}.
\end{multline*}
Using the same computation carried out from (\ref{W1}) to
(\ref{eq:16}), we get that \begin{equation}
\overline{\mathcal{N}}((E-E^{\alpha}))-C\mathbb{P}(\Omega_{n,E,\alpha})\leq
\mathcal{N}(E).
\end{equation}
Now we have to estimate the following probability,
$\mathbb{P}(\Omega_{n,E,\alpha})$. It is the purpose of the
following Lemma. It is a large deviation argument.
\begin{lem}\label{le:2}There exists $\tau>0$ such that for  $E$ sufficiently small and $n$ large,  we have
$$\mathbb{P}(\Omega_{n,E,\alpha})\leq e^{-E^{-\tau}}.$$
\end{lem}
Now the proof of Theorem \ref{thr:1} is just to take into account
Theorem \ref{thr:5}  and Lemma \ref{le:2}\newline {\bf{The proof
of Lemma \ref{le:2}}}\newline We prove this Lemma using techniques
of \cite{Kl:01b}. We have
$\Omega_{n,E,\alpha}\subset\Omega'_{n,E,\alpha}$.\newline
 With
\begin{multline*}
\Omega '_{n,E,\alpha}=\Big\{ \omega ;\ \exists  u\in
H^1(\mathbb{R}) ;\| u\|_{L^2(\mathbb{R}^d)}=1; \ \|\nabla u\|^2
\leq E\rho_* ;\\ \text{and}\ \Big|\langle
\Big(H_{\omega}^{n}(\theta)-\overline{H}^n(\theta)\Big) u,u\rangle
\Big|\geq E^{\alpha} \|u\|^2\ \Big\}.
\end{multline*}
Let us estimate the probability of the latest events. Notice that
we asked that
\begin{equation}
\Big|\langle
\Big(H_{\omega}^{n}(\theta)-\overline{H}^n(\theta)\Big) u,u\rangle
\Big|=\sum_{i=1}^d\Big|\langle
\Big(\rho_{\omega}^{n}(\theta)-\overline{\rho}^n(\theta)\Big)
\partial_{x_i}u,\partial_{x_{i}}u\rangle\Big |\geq E^{\alpha}
\|u\|^2.\label{W2}
\end{equation}
 Let $ u\in H^1(\mathbb{R}^d)$, then
$u$ can be written using the Floquet decomposition as:
$$
u=\sum_{k\in
\mathbb{N}}\int_{\mathbb{T}_{2n+1}^*}\chi_{k}(\theta)w_{k}(x,\theta)d\theta.
$$
Where $(w(\cdot,\theta)_k)_{k\in\mathbb{N}}$ are the Floquet
eigenfunctions of  $-\Delta_n^{\theta}$ associated to
$(E_k(\theta))_{k\in \mathbb{N}}$.\newline By this, for $u$ such
that $\langle -\Delta u,u\rangle \leq E\rho_*$ we have:
\begin{equation}\label{a}\Big(\sum_{k\geq0}\int_{\mathbb{T}^*_{2n+1}}|E_k(\theta)|^2|
\chi_{k}(\theta)|^2d\theta\Big)\leq C E^2.
\end{equation} $0$ is the bottom of the spectrum of
$-\Delta$. It is a simple non-degenerate Floquet eigenvalue
\cite{21}. Hence there exists $C>0$ such that
\begin{itemize}
\item For $k\neq 0, \forall \theta\in \mathbb{T}^*_{2n+1}$
\begin{equation}\label{b}
|E_k(\theta)|\geq 1/C,
\end{equation}
\item and $\exists Z=\{\theta _j\in \mathbb{T}^*_{2n+1};\ 1\leq j\leq
n_0\}$ such that $E_0(\theta_j)=0$.
\begin{equation}\label{c}
|E_0(\theta)|\geq 1/C \inf  _{1\leq j\leq n_0}|\theta-\theta_j|^2.
\end{equation}
\end{itemize}
Let $(2l+1)=[E^{-1/2+2\rho'}]_{\circ}\cdot[E^{-\rho'}]_{\circ}$
and $(2k+1)=[E^{-\eta}]_{\circ}$, where $\alpha
<\rho'<\frac{d}{4(d+1)}$  and $\eta>0 $ such that $(2l+1)\cdot
(2k+1)=2n+1$. Here $[\cdot]_{\circ}$ denotes the largest odd
integer smaller than $\cdot$.\newline From (\ref{a}), (\ref{b})
and (\ref{c}) we get that
\begin{equation}
\sum_{k\geq 1}\int_{\mathbb{T}^*_{2n+1}}|\chi_{k}(\theta)|^2
d\theta+\sum_{j=1}^{n_{0}}\int
_{|\theta-\theta_j|>\frac{1}{l}}|\chi_{0}(\theta)|^2 d\theta\leq
CE^{2}l^2\leq CE^{2\rho'}.
\end{equation}
Hence we write
\begin{equation}
u=\sum_{j=1}^{n_0}u_j+u^e,\ \ {\text{where}}\ \
u_j=\int_{|\theta-\theta_j|\leq
\frac{1}{l}}\chi_{0}(\theta)w_{0}(\cdot,\theta_j)d\theta;\ \
\|u^{e}\|\leq CE^{2\rho'}, \label{W3}
\end{equation}
and we have
$$
\sum_{j=1}^{n_{0}} \|u_j\|^2= \|u\|- CE^{\rho'}=1-CE^{2\rho'}.
$$
Now using (\ref{W3}) in (\ref{W2}), we get that for $E$ small we
have
\begin{equation}
\sum_{1\leq j,j'\leq
n_0}\Big|\langle\Big(\rho_{\omega}^{n}-\overline{\rho}^{n}\Big)\nabla
u_j,\nabla u_{j'}\rangle \Big|\geq E^{\alpha}/4.\label{Wparis}
\end{equation}
So, for some $1\leq j,j'\leq n_0$, one has
\begin{equation}
\Big |\langle\Big(\rho_{\omega}^{n}-\overline{\rho}^{n}\Big)\nabla
u_j,\nabla u_{j'}\rangle \Big | \geq
E^{\alpha}/(2n_{0})^2.\label{W2paris}
\end{equation}
Now we state a Lemma based on the Uncertainly principle and proved
in \cite{Kl:01b}.
\begin{lem}\cite{Kl:01b}Fix $1\leq j\leq n_{0}$. For $1\leq l'\leq l$ there exists $\widetilde
{u}_{j}\in L^{2}(\mathbb{R}^d)$ such that $\newline 1)\
\widetilde{u}_{j}$ is constant on each cube
$$
\Lambda_{\gamma,l'}=\{x=(x_1,\cdots,x_d);\ \forall 1\leq i \leq d
-l'-\frac{1}{2}\leq x_i-(2l'+1)\gamma_i<l'+\frac{1}{2}\}$$ where
$\displaystyle\gamma=(\gamma_1,\cdots,\gamma_d)\in \mathbb{Z}^d
.$\newline $2)\ \exists C>0$ depending only on $w_0(\cdot,\theta)$
such that
\begin{equation}\| u_j-\widetilde u_j\cdot
\overline{w}_0(\cdot,\theta_{j})\|_{L^2(\mathbb{R} ^d)}\leq Cl'/l,
\label{W5}
\end{equation}where $\overline{w}_0(\cdot,\theta)$ is the periodic
component of $w_0(\cdot,\theta)$ i.e
$w_0(\cdot,\theta)=e^{ix\theta}\overline{w}_0(\cdot,\theta)$.
\end{lem}
Let
$$\psi _j(x)=\widetilde {u}_j(x)\overline{w}_{0}(x,\theta_{j})=\overline{w}_0(x,\theta_{j})\sum_{\beta\in
\mathbb{Z}^d}(2l'+1)^{-d/2}a_j(\beta)\mathbf{1}_{(2l'+1)\beta
+\Lambda_{0,l'}}.$$ $\widetilde
u_j(x)\overline{w}_0(x,\theta_{j})\in L^{2}(\mathbb{R}^d)$ using
the periodicity of $\overline{w}_0(x,\theta_{j})$ we get,
\begin{multline}
\|\psi_{j}\|_{L^2(\mathbb{R}^d)}=\|\widetilde
{u}_j(x)\overline{w}_0(x,\theta_{j})\|_{L^{2}(\mathbb{R}^d)}=\\
\sum_{\beta\in\mathbb{Z}^d}|a_j(\beta)|^2\int_{\Lambda_{0,0}}|\overline{w}_0(\cdot,\theta_{j})|^{2}dx.
\end{multline} Then using (\ref{W5}) and the fact that
$\displaystyle
\int_{\Lambda_{0,0}}|w_{0}(x,\theta_{j})|^2dx=\int_{\Lambda_{0,0}}|\overline{w}_0(x,\theta_{j})|^2dx;
$ we get that there exists $C>0$ such that
\begin{equation}\label{wa}
\sum_{\beta\in \mathbb{Z}^d}|a_j(\beta)|^2\leq C
\|u_j\|_{L^2(\mathbb{R} ^d)}<+\infty.
\end{equation}
We set $2l'+1=[E^{-1/2+2\rho'}]_{\circ}$ and $2k'+1=
[E^{-\rho'}]_{\circ}\cdot[E^{-\eta}]_{\circ}$, for
$\alpha<\rho'<\frac{d}{4(d+1)}$ and $\eta>0$ so that
$(2n+1)=(2l'+1)\cdot(2k'+1)$. So taking into account
(\ref{W2paris}), (\ref{W5}) and the choice of $l$ and $l'$, we get
\begin{equation}
\sum_{i=1}^{d}
\Big|\langle(\rho_{\omega}^{0,n}-\overline{\rho}_{\omega}^{0,n})
\partial_{x_{i}}\psi_j, \partial_{x_{i}} \psi_{j'}\rangle \Big|\geq
E^{\alpha}/(2n_{0})^2-CE^{\rho'}\geq E^{\alpha}/(4n_{0})^2.
\end{equation}
We set
$$
\sum_{i=1}^d\Big|\langle(\rho_{\omega}^{0,n}-\overline{\rho^{0,n}})
\partial_{x_{i}}\psi_{j}, \partial_{x_{i}}\psi_{j'}\rangle \Big |=\sum_{1\leq i\leq d}|A^i_{j,j'}|.$$
With
$$
A^{i}_{j,j'}=\langle(\rho_{\omega}^{0,n}-\overline{\rho^{0,n}})
\partial_{x_{i}}\psi_{j}, \partial_{x_{i}}\psi_{j'}\rangle .
$$
We have
\begin{multline*}A^{i}_{j,j'}=\sum_{
\beta\in\mathbb{Z}^d}\sum_{\gamma\in\mathbb{Z}_{2n+1}^d}(2l'+1)^{-d}a_j(\beta)\cdot \overline{a_{j'}}(\beta) \cdot\\
\cdot
\int_{(2l'+1)\beta+\Lambda_{0,l'}}(\rho_{\omega}^{0,n}-\overline{\rho^{0,n}})(x-\gamma)\partial_{x_{i}}\overline{w}_0(x,\theta_j)\cdot
\overline{\partial_{x_{i}}\overline{w}_0(x,\theta_j')} dx
\end{multline*}
\begin{multline}= \sum_{\beta\in\mathbb{Z} ^d}\sum_{\gamma \in \mathbb{Z}^d_{2n+1}}
a_{j}(\beta)\cdot\overline{a_{j'}(\beta)}\cdot\\
\frac{1}{(2l'+1)^d}
\int_{\Lambda_{0,l'}}(\rho_{\omega}^{0,n}-\overline{\rho^{0,n}})(x-\gamma+(2l'+1)\beta)\partial_{x_{i}}\overline{w}_0(x,\theta_{j})\cdot\overline{\partial_{x_{i}}\overline{w}_0(x,\theta_{j'})}dx
.\end{multline} As $\rho_{\omega} ^n$ is
$(2n+1)\mathbb{Z}^d$-periodic and $(2l'+1)(2k'+1)=(2n+1)$ we get
that\begin{multline}A^{i}_{j,j'}= \sum_{\beta\in\mathbb{Z}
^d_{2k'+1}}\sum_{\beta'\in\mathbb{Z} ^d}\sum_{\gamma \in
\mathbb{Z}^d_{2n+1}}
a_{j}(\beta+(2k'+1)\beta')\cdot\overline{a_{j'}(\beta+(2k'+1)\beta')}\cdot\\
\frac{1}{(2l'+1)^d}
\int_{\Lambda_{0,l'}}(\rho_{\omega}^{0,n}-\overline{\rho^{0,n}})(x-\gamma+(2l'+1)\beta)
\partial_{x_{i}}\overline{w}_0(x,\theta_{j})\overline{\partial_{x_{i}}\overline{w}_0(x,\theta_{j'})}dx
.\end{multline} Using the expression of $\rho_\omega$ we get that.
\begin{multline}A^{i}_{j,j'}= \sum_{\beta\in\mathbb{Z}
^d_{2k'+1}}\sum_{\gamma \in
\mathbb{Z}^d_{2n+1}}\sum_{\beta'\in\mathbb{Z}^d}(\omega_\gamma-\overline{\omega})a_{j}(\beta+(2k'+1)\beta')\cdot\overline{a_{j'}(\beta+(2k'+1)\beta')}\cdot\\
\frac{1}{(2l'+1)^d}
\int_{\Lambda_{0,l'}}\rho^{0}(x-\gamma+(2l'+1)\beta)\partial_{x_{i}}\overline{w}_0(x,\theta_{j})\cdot\overline{\partial_{x_{i}}\overline{w}(x,\theta_{j'})}dx
.\end{multline} We set
\begin{equation}
B^{i}_{j,j'}(\beta)=\sum_{\beta'\in \mathbb{Z}^d}
a_{j}(\beta+(2k'+1)\beta')\cdot\overline{a_{j'}(\beta+(2k'+1)\beta')}.
\end{equation}
Then we get that \begin{multline*}
A^{i}_{j,j'}=(2l'+1)^{-d}\sum_{\gamma\in
\mathbb{Z}^d_{2n+1}}(\omega_\gamma-\overline{\omega})
\Big(\sum_{\beta\in\mathbb{Z}_{2k'+1}}B^{i}_{j,j'}(\beta)\cdot\\
\int_{\Lambda_{0,l'}}\rho^{0}(x-\gamma+(2l'+1)\beta)\overline{w}_0(x,\theta_{j})\cdot\overline{\overline{w}(x,\theta_{j'})}dx\Big)
\end{multline*}
\begin{multline}
=(2l'+1)^{-d}\sum_{\gamma\in
\mathbb{Z}^d_{2l'+1}}\Big[\sum_{\gamma'\in\mathbb{Z}^d_{2k'+1}}
(\omega_{\gamma+(2l'+1)\gamma'}-\overline{\omega})\Big(\sum_{\beta\in\mathbb{Z}_{2k'+1}}B^{i}_{j,j'}(\beta)\cdot\\
\int_{\Lambda_{0,l'}}\rho^{0}(x-\gamma+(2l'+1)(\beta-\gamma'))
\partial_{x_{i}}\overline{w}_0(x,\theta_{j})\cdot\overline{\partial_{x_{i}}\overline{w}(x,\theta_{j'})}dx\Big
)\Big].
\end{multline}
We set
$$C^{i}_{j,j'}(\gamma,\gamma')= \sum_{\beta\in\mathbb{Z}^d_{2k'+1}}B^{i}_{j,j'}(\beta)\int_{\Lambda_{0,l'}}\rho^{0}
(x-\gamma+(2l'+1)(\beta-\gamma'))\partial_{x_{i}}\overline{w}_0(x,\theta_{j})\cdot\overline{\partial_{x_{i}}\overline{w}_0(x,\theta_{j'})}dx$$
and
$$ Y^{i}_{j,j'}(\gamma)=\sum_{\gamma'\in\mathbb{Z}^d_{2k'+1}}
(\omega_{\gamma+(2l'+1)\gamma'}-\overline{\omega})C^{i}_{j,j'}(\gamma,\gamma').$$
Then \begin{equation}
A_{j,j'}^{i}=\frac{1}{(2l'+1)^d}\sum_{\gamma\in
\mathbb{Z}^d_{2l'+1}}Y^{i}_{j,j'}(\gamma).
\end{equation}
Notice that $(Y^{i}_{j,j'}(\gamma))_{\gamma\in
\mathbb{Z}^{d}_{2l'+1}}$ are bounded random and independent
variables with $\mathbb{E}(Y^{i}_{j,j'}(\gamma))=0$. Indeed, using
the fact that $\rho^{0}$ is compactly supported and (\ref{wa}) we
get that
$$
|Y_{j,j'}(\gamma)|\leq ||u^0||^2.
$$
So, to estimate the probability of $\Omega(n,E,\alpha)$ it
suffices to estimate the probability that
\begin{equation}
E^{\alpha}/(4n_{0})^{2}\leq \frac{1}{(2l'+1)^d}\sum_{\gamma\in
\mathbb{Z}^d_{2l'+1}}Y^{i}_{j,j'}(\gamma).
\end{equation}
This probability is given by the the large deviation principle
which gives that \cite{De-Ze:92}
$$
\mathbb{P}\Big( E^{\alpha}/(4n_{0})^{2}\leq
\frac{1}{(2l'+1)^d}\cdot\sum_{\gamma\in
\mathbb{Z}^d_{2l'+1}}Y^{i}_{j,j'}(\gamma)\Big)\leq
e^{-c(l')^{d}E^{2\alpha}}\leq e^{-cE^{-d/2+2d\rho'+2\alpha}}.
$$
Here we have used the expression of $l'$. Using the fact that for
our choice of $\rho'$ we have $-d/2+2d\rho'+2\alpha<0$, so for
some $\tau>0$ and $E$ sufficiently small, one has
$$
\mathbb{P}\Big( E^{\eta'}\leq
\frac{1}{(2l'+1)^d}\cdot\sum_{\gamma\in
\mathbb{Z}^d_{2l'+1}}Y^{i}_{j,j'}(\gamma)\Big)\leq
e^{-E^{-\alpha}}.
$$
As the probability of $\Omega(n,E,\alpha)$ is bounded by the sum
over $1\leq i\leq d$ and $1\leq j,j'\leq n_{0}$ of the probability
estimate previously, we get the result of the Lemma \ref{le:2}
 $\Box$
\bibliographystyle{plain}
\bibliography{biblio}

\end{document}